\documentclass{ws-ijmpe}
\usepackage[super,compress]{cite}
\begin{document}

\markboth{G. Nikoghosyan et al.}{Isovector pair correlations in analytically solvable models}

\catchline{}{}{}{}{}

\title{Isovector pair correlations in analytically solvable models}

\author{G. Nikoghosyan, A. Balabekyan}
\address{Yerevan State University, 375049 Yerevan,  Armenia}

\author{E.A. Kolganova, R.V. Jolos\footnote{Corresponding author}}
\address{Bogoliubov Laboratory of Nuclear Physics\\
Joint Institute for Nuclear Research, 141980 Dubna, Russia,\\
Dubna State University, 141982 Dubna, Russia\\
*jolos@theor.jinr.ru}
\author{D.A. Sazonov}
\address{Dubna State University, 141982 Dubna, Russia}

\maketitle

\begin{history}
\end{history}

\begin{abstract}
The eigensolutions of the collective Hamiltonian with different potentials suggested
for description of the isovector pair correlations are obtained, analyzed and compared
with the experimental energies. It is shown that the isovector pair correlations in nuclei around 
$^{56}$Ni
can be described as anharmonic pairing vibrations. The results obtained indicate the presence
of the $\alpha$-particle type correlations in these nuclei and the existence of the interaction different from isovector pairing which also influences on the isospin dependence of the energies.
\end{abstract}


\ccode{PACS numbers: 21.10.-k, 21.10.Dr, 21.60.Ev}


\section{Introduction}

Pair correlations of nucleons in atomic nuclei play an important role
in description of their properties \cite{Bohr1,Belyaev,Soloviev,Zel}. In heavy nuclei where
protons and neutrons occupy single particle states belonging to
different shells, pair correlations are considered 
as acting among nucleons of the same kind. In nuclei close to
the magic ones static limit of the pair correlations is not achieved.
In this case ground states of the neighboring  nuclei form pairing
vibrational bands \cite{Bohr2}. With increase of the number of valence
nucleons the limit of the static pair correlations is realized. In this
limit ground states of the neighboring even-even nuclei form pairing
rotational bands  \cite{Bohr2}. There is a transitional region between
these two limits where the phase transition from one type of pairing
excitations to the other one takes place \cite{Macchiavelli1}.

In the case of the relatively light nuclei with $N\approx Z$ it is
necessary to consider not only neutron-neutron and proton-proton
but also neutron-proton pair correlations. In other words, isospin quantum
number should be taken into account. In principle, both isovector $T=1$
and isoscalar $T=0$ pair correlations can take place. However, since
there are no clear indications on the presence of the isoscalar pair
correlations (see, however, \cite{Frauendorf,Sagawa,Gerzelis}) we consider below only isovector pairing and their
consequences.

Just after the introduction of the concept of the collective pair excitations
the collective Hamiltonian has been constructed to treat the isovector pair excitations.
In \cite{Dussel1} it was done by analogy with the Bohr
collective Hamiltonian for description of the quadrupole collective motion.
In \cite{Jolos1} it was done by treating isovector pair correlations applying
boson representation technique. 

The aim of the present paper is to obtain and analyze the solutions of the
collective Hamiltonian for isovector pairing obtained with different
collective potentials which include all physical situations from pairing
vibrational to pairing rotational limits and allow to consider a
transitional region. The results of calculations  allow to make
a conclusion about the character of the collective pairing motion in
these nuclei.

\section{Collective Hamiltonian for description of the isovector monopole pairing}

There are two collective modes, namely, pairing addition and pairing
removal modes needed to describe pair correlations in nuclei \cite{Bohr2}.
Each of them is characterized by three projections of isospin corresponding
to neutron-neutron, proton-proton and neutron-proton correlated pairs.
Thus, in total there are six collective variables. They can be presented
by the complex isovector $z_{1 \mu}(\mu=0,\pm 1)$. It was suggested in \cite{Dussel1}
to separate in $z_{1\mu}$ the variables related to isospin invariance and gauge
invariance
\begin{eqnarray}
\label{eq1}
z^+_{1\mu}=\Delta\exp(-\imath\phi)\left(D^1_{\mu 0}(\psi_1,\psi_2,\psi_3)\cos\theta \phantom{\frac{1}{\sqrt{2}}}\right.\nonumber\\
\left.+\frac{1}{\sqrt{2}}(D^1_{\mu 1}(\psi_1,\psi_2,\psi_3)+D^1_{\mu -1}(\psi_1,\psi_2,\psi_3))\sin\theta\right).
\end{eqnarray}
Here $D^1_{\mu 1}(\psi_k,\psi_2,\psi_3)$ is the Wigner function and $\psi_1,\psi_2,\psi_3$
are Euler angles in isospace. The angle $\phi$ is conjugate to the
operator of the number of the nucleon pairs ${\hat N}=-\imath\partial/\partial\phi$
added or removed from the basic nucleus. The variable $\Delta$ characterizes
a strength of the pair correlations and $\theta$ describes isospin structure
of the pair correlations in the intrinsic frame. The collective coordinates
can be presented also in terms of the boson creation and annihilation operators
\begin{eqnarray}
\label{eq2}
z^+_{1\mu}=\frac{1}{\sqrt{2}}\left(\beta^+_{1\mu}(+)+(-1)^{1-\mu}\beta_{1-\mu}(-)\right),
\end{eqnarray}
where $\beta^+_{1\mu}(+)$ and $\beta_{1-\mu}(-)$ correspond to addition and removal modes, respectively.

Written in terms of the collective coordinates the collective Hamiltonian takes the form
\cite{Dussel1,Jolos1}
\begin{eqnarray}
\label{eq3}
{\hat H}=\frac{1}{2B}\left(-\frac{1}{\Delta^5}\frac{\partial}{\partial\Delta}\Delta^5\frac{\partial}{\partial\Delta}-\frac{1}{\Delta^2\sin 4\theta}\frac{\partial}{\partial\theta}\sin 4\theta\frac{\partial}{\partial\theta}\right.\nonumber\\
+\left.\frac{1}{\Delta^2}\left(\frac{{\hat T}^2_x}{\cos^2 2\theta}+\frac{{\hat T}^2_y}{\cos^2 \theta}+\frac{{\hat T}^2_z}{\sin^2 \theta}+\frac{2\sin2\theta}{\cos^2 2\theta}{\hat N}{\hat T}_x+\frac{{\hat N}^2}{\cos^2 2\theta}\right)\right)\nonumber\\
+V(\Delta^2, \Delta^4\cos^2 2\theta).
\end{eqnarray}
Here ${\hat T}_{x,y,z}$ is isospin projection operator. It is shown in \cite{Dussel1} that
the angles $\theta$ and $\phi$ vary in the limits $0\le\theta\le\frac{\pi}{4}$ and $0\le\phi\le\pi$.
The volume element is
\begin{eqnarray}
\label{eq4}
d\tau=\frac{1}{4}\Delta^5|\sin 4\theta |d\Omega d\phi d\theta d\Delta .
\end{eqnarray}
The operator
\begin{eqnarray}
\label{eq5}
{\hat \Lambda}^2=-\frac{1}{\sin 4\theta}\frac{\partial}{\partial\theta}\sin 4\theta\frac{\partial}{\partial\theta}
+\frac{{\hat T}^2_x}{\cos^2 2\theta}+\frac{{\hat T}^2_y}{\cos^2 \theta}+\frac{{\hat T}^2_z}{\sin^2 \theta}+\frac{2\sin2\theta}{\cos^2 2\theta}{\hat N}{\hat T}_x+\frac{{\hat N}^2}{\cos^2 2\theta},
\end{eqnarray}
depends only on the angle variables and is an analog of the corresponding
operator of the Bohr Hamiltonian for the quadrupole mode \cite{Rowe1}. Its eigenfunctions are
\begin{eqnarray}
\label{eq6}
{\hat \Lambda}^2\Phi^{T,N,\lambda}_{T_z}(\theta,\phi, {\vec \psi})=\lambda (\lambda +4)\Phi^{T,N,\lambda}_{T_z}(\theta,\phi, {\vec \psi}),
\end{eqnarray}
where $\lambda$ is the number of the added and removed bosons which are not coupled to zero isospin. Thus, $\lambda$ in the same nucleus is changed by two units.

It was shown in \cite{Jolos1} that the collective potential can be presented as
\begin{eqnarray}
\label{eq7}
V=V_0(\Delta^2)+A\Delta^4(1-\frac{1}{2}\cos^2 2\theta),
\end{eqnarray}
where $A$ is positive. Thus the potential has a minimum at $\theta$=0.

Below, following the analogy with consideration of critical symmetries in
nuclear structure physics related to collective quadrupole excitations \cite{Iachello1,Iachello2,Iachello3}
we will consider only those cases where dynamic variables $\Delta$ and $\theta$
can be separated in the collective Hamiltonian.

We will consider below two cases distinguished by the character of a $\theta$-dependence of the
potential. In the first case, we assume that potential has a deep minimum at $\theta$=0.
In the second case, we assume that potential does not depend on $\theta$. Concerning $\Delta$-dependence we will consider two limiting cases: harmonic oscillator and a potential with very deep and narrow minimum at $\Delta=\Delta_0\ne 0$, and additionally two intermediate cases: square well potential and Davidson potential  \cite{Davidson,Bonatsos}.

\section{Analytically solvable models}

\subsection{Harmonic oscillator}

In this case $V(\Delta^2)=\frac{1}{2}C\Delta^2$. The total wave function is factorized in two multipliers
\begin{eqnarray}
\label{eq9}
\Psi=\Phi^{TN\lambda}_{T_z}(\theta,\phi, {\vec \psi})g_{n\lambda}(\Delta^2),
\end{eqnarray}
where $\Phi^{TN\lambda}_{T_z}$ is introduced in (\ref{eq6}) and $g_{n\lambda}(\Delta^2)$
is a solution of the following equation
\begin{eqnarray}
\label{eq10}
\left(\frac{1}{2B}\left(-\frac{1}{\Delta^5}\frac{\partial}{\partial\Delta}\Delta^5\frac{\partial}{\partial\Delta}+\frac{\lambda(\lambda+4)}{\Delta^2}\right)+\frac{1}{2}C\Delta^2\right)g_{n\lambda}(\Delta^2)=Eg_{n\lambda}(\Delta^2)
\end{eqnarray}
Thus,
\begin{eqnarray}
\label{eq11}
g_{n\lambda}(\Delta^2)={\cal N}_{n\lambda}\exp\left(-\frac{1}{2}\sqrt{BC}\Delta^2\right)\left((BC)^{1/4}\Delta\right)^{\lambda}L_n^{(\lambda+2)}(\sqrt{BC}\Delta^2)
\end{eqnarray}
and
\begin{eqnarray}
\label{eq12}
E=\sqrt{\frac{C}{B}}(2n+\lambda+3)
\end{eqnarray}
The ground state of the basic nucleus is determined by $n$=0 and $\lambda$=0.
The first excited multiplet consists in the states with $n$=0, $\lambda=2$
and $n$=1, $\lambda$=0. The ground state of the nucleus with two nucleons more than in the basic
one is characterized by $n$=0, $\lambda$=1 and so on.

\subsection{Limit of the pairing and isospin rotations. Static pair correlations.}

If the potential energy has a deep and narrow minimum at $\theta$=0 and $\Delta=\Delta_0\ne 0$,
the wave function of the low-energy pairing- and isospin - rotational states
take the form
\begin{eqnarray}
\label{eq13}
\Psi_{NTT_z}=\sqrt{\frac{2T+1}{8\pi^2}}\exp(\imath N\phi)D^T_{T_z 0}({\vec \psi})\exp\left(-\frac{1}{2}\frac{\theta^2}{\theta_{00}^2}\right)\exp\left(-\frac{1}{2}\frac{(\Delta-\Delta_0)^2}{\Delta_{00}^2}\right),
\end{eqnarray}
and the excitation energies are
\begin{eqnarray}
\label{eq14}
E_{NT}=\frac{T(T+1)+N^2}{2B\Delta_0^2}.
\end{eqnarray}
 The excited states corresponding to the vibrations of $\theta$ and $\Delta$ around their equilibrium values have higher excitation energies.

\subsection{Infinitely deep square well potential}

We assume that the collective potential as a function of $\theta$ has a deep
and narrow minimum at $\theta$=0 and $V(\Delta)$ has a form of the square well potential
\begin{eqnarray}
\label{eq15}
V(\Delta)=
\begin{cases}
0, \Delta \le \Delta_l \\
\infty, \Delta > \Delta_l
\end{cases}
\end{eqnarray}
We consider below only those cases when there are no excitations of $\theta$-vibrations.
Then the $\Delta$-dependent part of the total wave function satisfies to the
following equation
\begin{eqnarray}
\label{eq16}
\left(-\frac{1}{2B}\frac{1}{\Delta^5}\frac{\partial}{\partial\Delta}\Delta^5\frac{\partial}{\partial\Delta}+\frac{T(T+1)+N^2}{2B\Delta^2}\right)\psi(\Delta)=E\psi(\Delta)
\end{eqnarray}
with boundary condition $\psi(\Delta_l)$=0. The solution of the
eigenvalue problem is the following:
\begin{eqnarray}
\label{eq17}
\psi(\Delta)=\frac{1}{\Delta^2}J_{\nu}\left(x_{nTN}\frac{\Delta}{\Delta_l}\right),\nonumber\\
E=\frac{x_{nTN}^2}{2B\Delta^2_l}.
\end{eqnarray}
Here $J_{\nu}$ is Bessel function, $\nu=\sqrt{T(T+1)+N^2+4}$, $x_{nTN}$
is the zeroes of the Bessel function, and index $n$ numbers the zeroes of the Bessel
function.

In the case when potential does not depend on $\theta$ and its $\Delta$-dependence is described by (\ref{eq15})
the $\Delta$-dependent part of the total wave function satisfies to the following equation
\begin{eqnarray}
\label{eq18}
\left(-\frac{1}{2B}\frac{1}{\Delta^5}\frac{\partial}{\partial\Delta}\Delta^5\frac{\partial}{\partial\Delta}+\frac{\lambda(\lambda+4)}{2B\Delta^2}\right)\psi(\Delta)=E\psi(\Delta),
\end{eqnarray}
with the former boundary condition $\psi(\Delta_l)$=0. Then
\begin{eqnarray}
\label{eq19}
\psi(\Delta)=\frac{1}{\Delta^2}J_{\nu}(x_{n\lambda}\frac{\Delta}{\Delta_l})
\end{eqnarray}
and
\begin{eqnarray}
\label{eq20}
E=x^2_{n\lambda}\frac{1}{2B\Delta^2_l}.
\end{eqnarray}
In (\ref{eq17}) $\nu=\lambda+2$ and $x_{n\lambda}$ is  zeroes of the Bessel function.

\subsection{Davidson potential}

In this subsection we consider a case when $\Delta$-dependence of the potential is  taken in the form of
Davidson potential:
\begin{eqnarray}
\label{eq21}
V(\Delta)=\frac{1}{2}C\left(\frac{\Delta_0^4}{\Delta^2}+\Delta^2\right)
\end{eqnarray}
In the case when  potential as a function of $\theta$ has a very deep and narrow minimum,
the $\Delta$-dependent part of the total wave function satisfies to the following
equation
\begin{eqnarray}
\label{eq22}
\left(-\frac{1}{2B}\frac{1}{\Delta^5}\frac{\partial}{\partial\Delta}\Delta^5\frac{\partial}{\partial\Delta}+\frac{T(T+1)+N^2+BC\Delta_0^4}{2B\Delta^2}\right)\psi(\Delta)=E\psi(\Delta),
\end{eqnarray}
The solution of the eigenvalue problem is the following
\begin{eqnarray}
\label{eq23}
\psi(\Delta)=\exp\left(-\frac{1}{2}\sqrt{BC}\Delta^2\right)\left((BC)^{1/4}\Delta\right)^{\alpha}L^{\alpha+2}_n\left(\sqrt{BC}\Delta^2\right),\nonumber\\
E_{T,N,n}=\sqrt{\frac{C}{B}} \left( 2n+\sqrt{T(T+1)+N^2+\rho^4_0+4}+1 \right),
\end{eqnarray}
where
\begin{eqnarray}
\label{eq24}
\rho_0=(BC)^{1/4}\Delta_0,\nonumber\\
\alpha=\sqrt{T(T+1)+N^2+\rho^4_0+4}-2.
\end{eqnarray}
Varying $\rho_0$ we can investigate a transition region between
pairing vibrational and pairing rotational limits. In the limit
of very large $\rho_0$ we obtain from (\ref{eq24}) that
\begin{eqnarray}
\label{eq25}
E\approx\sqrt{\frac{C}{B}}(2n+1)+C\Delta_0^2+\frac{2}{B\Delta_0^2}+\frac{T(T+1)+N^2}{2B\Delta_0^2}.
\end{eqnarray}
The excitation energies are given by the expression
\begin{eqnarray}
\label{eq26}
E^*=\frac{T(T+1)+N^2}{2B\Delta_0^2}+2\sqrt{\frac{C}{B}}n.
\end{eqnarray}
Thus, the low-lying states are the pairing and isospin rotational excitations and
(\ref{eq26}) coincides with the result given in (\ref{eq14}).

If we consider a case of $\theta$-independent potential we obtain the following
expression for the energy
\begin{eqnarray}
\label{eq27}
E_{\lambda,n}=\frac{C}{B}(2n+\sqrt{(\lambda+2)^2+\rho_0^4}+1).
\end{eqnarray}
which coincide with the result obtained in (\ref{eq12}) in the limit
of $\rho_0=0$.

\section{Comparison of the model results with the experimental data}

To compare the results of calculations with the experimental data the experimental energies have to be reduced to quantities which can be directly compared
with the model results. For this, we subtract from the empirical binding energies those
contributions that are generated by the sources other than the isovector monopole pair
correlations. For nuclei around the basic nucleus with $A=A_0$ and $Z=Z_0$ we define
the quantity
\begin{eqnarray}
\label{eq28}
E(A,Z)=-\left(B_{exp}(A,Z)-B_{LD}(A,Z)\right)+\left(B_{exp}(A_0,Z_0)-B_{LD}(A_0,Z_0)\right)
\end{eqnarray}
where $B_{exp}(A,Z)$ is the experimental binding energy of the nucleus with mass number $A$ and charge $Z$. The quantity
$B_{LD}(A,Z)$ is defined by the liquid drop mass formula without symmetry energy and the pairing energy terms
\begin{eqnarray}
\label{eq29}
B_{LD}(A,Z)=a_vA-a_sA^{2/3}-a_c\frac{Z(Z-1)}{A^{1/3}},
\end{eqnarray}
where $a_v$=15.75 MeV, $a_s$=17.8 MeV, $a_c$=0.711 MeV.
The symmetry energy term which can be presented as $-4a_AT(T+1)/A$
contains dependence of the ground state energy on isospin. However,
dependence of the nuclear binding energies on isospin is introduced also by the isovector pair
correlations. This is the reason why we do not include in $B_{LD}$ the symmetry
energy term in order to see what part of the isospin dependence of the binding energy
is contained in the isovector pair correlations. We do not subtract also the pairing energy
term, which is usually presented as $a_p/A^{1/2}$, since we expect that isovector pairing forces reproduce
this effect.

The results obtained in calculations with different choices of the potential
 are compared below with the experimental data for nuclei around $^{56}$Ni.
Thus, $^{56}$Ni is our basic nucleus.

The results of calculations of the relative energies, i.e. $(E(A,T)-E(A=56,T=0))$, of the ground states of nuclei with the values of isospin from $T$=0 to $T$=4 and the experimental data are shown in Fig.1. The given results include only energies of the even-even
nuclei with $A\ge 56$. The results of calculations are shown for the following variants
of the collective potentials: potential with a deep minimum at $\theta$=0
and $\Delta$-dependence described by the square well potential and  potential with a deep minimum at $\theta$=0
and $\Delta$-dependence described by Davidson potential with the parameter $\rho_0$=1. For each calculation variant and the
experimental data energies are given in units $\left(E(N=1,T=1)-E(N=0,T=0)\right)$,
where $N=(A-56)/2$ is a number of the nucleon pairs added to the basic nucleus.

It is seen from Fig.1 that the results of the model calculations deviate the most from the
experimental data for the states with $T$=0. Moreover, this deviation increases
 with increasing  $N$. Probably, this indicates on the absence of the
 $\alpha$-particle type correlations \cite{Alpha,Sandulescu1,Sandulescu2,Sandulescu3} in the model Hamiltonian.
 The value of $\rho_0$ in the case of calculation with Davidson potential is selected so as to achieve, if possible, a better description of the
experimental data. In general, the results of calculations with Davidson
potential at $\rho_0$=1 are closer to the experimental data than calculations with the other
potentials. However, deviations from the experimental data are noticeable.
As it is seen in Fig.1, the energies of the states with $T$=1-4 are weakly
dependent on $N$, while the calculated energies of these states increase with $N$.

\begin{figure}[th]
\centerline{\includegraphics[width=7.3cm]{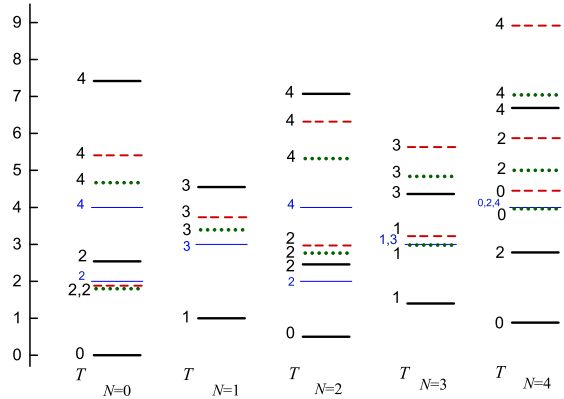}}
\caption{Experimental and calculated relative energies of the ground states 
of even-even nuclei with different values of the isospin $T$: thick solid line
(black) - experimental data, dashed line (red) - square well potential, dotted
line (green) - Davidson potential with $\rho_0$=1, thin solid line (blue) - harmonic oscillator potential. 
Energies are given in units $\left(E(N=1,N=1)-E(N=0,T=0)\right)$.}
\label{Fig:1}
\end{figure}

The results presented in Fig.1 include dependence of energies on both the mass
number and isospin at a fixed mass number. The results for the energies of states with $T=N$ obtained under
different assumptions on the potential are presented in Fig.2.  As shown in Fig.2, the
calculations with Davidson potential at $\rho_0$=1 are well consistent with the experimental ones.
Note, that both experimental and calculated energies shown in Fig.2 increase
with $N$ and $T$ much slower than it should be in the rotational limit
for both isospin and pairing rotations. In the case of Davidson potential,
such a limit is reached at $\rho_0\gg 1$. In this case, the energies of the states shown in Fig.2
are described by the following expression
\begin{eqnarray}
\label{eq29}
E(N,T)=\frac{1}{3}\left(T(T+1)+N^2\right)
\end{eqnarray}

\begin{figure}[th]
\centerline{\includegraphics[width=7.3cm]{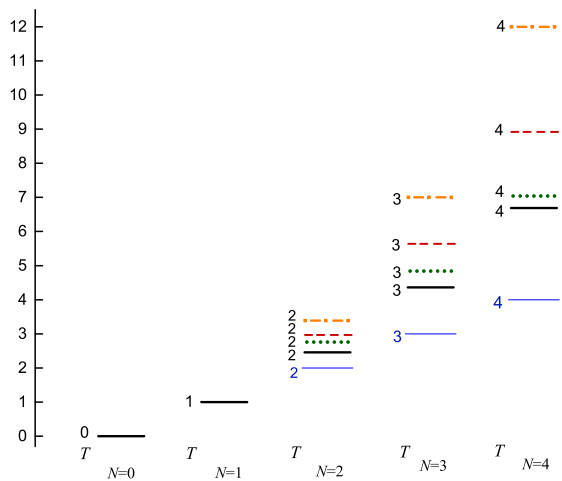}}
\caption{Experimental and calculated relative energies of the ground states
of even-even nuclei with isospin $T=N$. Thick solid line
(black) - experimental data, dashed line (red) - square well potential, dotted
line (green) - Davidson potential with $\rho_0$=1, dot-dashed line (orange) - Davidson 
potential with $\rho_0\rightarrow\infty$, thin solid line (blue) - harmonic oscillator potential.  
Energies are given in units $\left(E(N=1,N=1)-E(N=0,T=0)\right)$.}
\label{Fig:2}
\end{figure}

To separate the effects associated with a poor description of the energies
of the lowest states with $T=0$ the energies of states with  $T=1-4$ shown in Fig.3 
are calculated from the energy of the state with $T=0$. Only the states of the even-even nuclei are presented.
It can be seen from Fig.3 that both the experimental and calculated energies of the states with $T=1-4$
gradually decrease with the grows of $N$. At the same time, the calculated energies qualitatively
reproduce  dependence on $N$ of the experimental energies. Being counted from the energies of the states
with $T=0$, the calculated energies are smaller than the experimental ones.
This indicates that the moment of inertia for the isospin rotations in the model
Hamiltonian is significantly larger than the experimental one. Apparently,
this reflects some effects not taken into account in the model with isovector pairing.

\begin{figure}[th]
\centerline{\includegraphics[width=7.3cm]{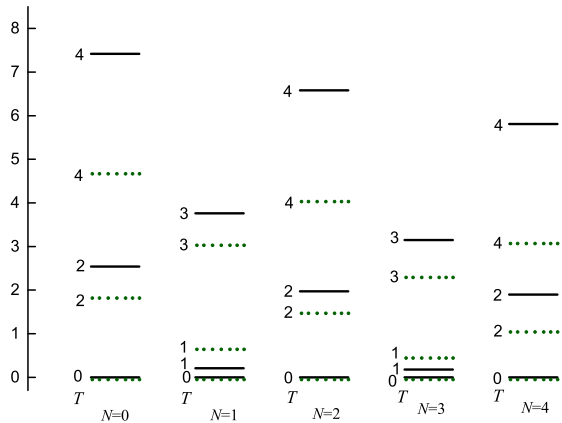}}
\caption{Experimental and calculated relative energies of the ground states
of even-even nuclei with different values of the isospin $T$. At all values of $N$
 energies are calculated from the energy of the state with $T=0$.  Thick solid line
(black) - experimental data,  dotted
line (green) - Davidson potential with $\rho_0$=1. 
Energies are given in units $\left(E(N=1,T=1)-E(N=0,T=0)\right)$.}
\label{Fig:3}
\end{figure}

Another effect is illustrated in Fig.4, where along with the energies of the
even-even nuclei the energies of the odd-odd ones are presented. The experimental energies
and the energies calculated with the Hamiltonian (\ref{eq3}) having a potential with deep and narrow minimum at $\theta$=0
whose $\Delta$-dependence is described by Davidson potential with $\rho_0=1$ are shown.
In Fig.4, just as in Fig.3, energies are counted from the energy of the state with $T=0$ with the same $N$.
As it is seen from (\ref{eq23}) the eigenvalues of the Hamiltonian can be found
for any set of $N$ and $T$. 
It can be seen from Fig.4 that the experimental and calculated energies of the states with even $T$,
which include states with $T=0$, whose energies are fixed at zero,
smoothly vary with $N$, while the experimental energies of the states with odd $T$ show
staggering. This irregularity in the isospin dependence of the experimental energies is also seen in the energy spectra at each value of $N$. 
In contrast to the behavior of the experimental energies, the calculated energies of the states with both even and odd $T$ vary smoothly with $N$.

Consider this irregularity in details. As it is seen in Fig.4 at $N=0$, states with $T=1$ and 2 are shifted in energy closer to each other forming a kind of a splitted multiplet. The states with $T=3$ and 4 are also shifted closer to each other but more separated from the states with $T=1$ and 2. Looking at the experimental spectra at $N=1$ we see a different picture. The state with $T=1$ is quite lower in energy compared to the states with $T=2$ and 3. States with $T=2$ and 3 are shifted closer to each other also forming a kind of a splitted multiplet. At the same time these states are quite separated in energy from the states with $T=4$. The situation at $N=2$ is similar to that at $N=0$, and the situation at $N=3$ is similar to that at $N=1$.

This analysis leads us to the following interpretation of the staggering phenomenon demonstrated in Fig.4. The assumption of the presence of a deep minimum at $\theta$=0  in the collective potential used in our calculations corresponds to the picture of a rigid isospin rotations with its rotational-like dependence of the energies on isospin which is far from the picture given by the harmonic oscillator. Although, the collective potential 
dependence on $\Delta$ described by Davidson potential with $\rho_0=1$ makes the energy dependence on isospin slightly different from the rotational one. As a result the calculated energies don't demonstrate the staggering effect.

The experimental spectra are rather close to the case of an anharmonic vibrator which qualitatively preserve the picture of the slightly splitted multiplets characteristic for the harmonic oscillator.
Indeed, in the case of the harmonic oscillator  and $N=0$ the states with $T=1$ and 2 belong to the same multiplet, but the states with $T=3$ and 4 belong to the other multiplet  with bigger energy.
In the case of the harmonic oscillator and $N=1$ the states with $T=2$ and 3 belong to the same multiplet and so on.

Thus, the experimental data on the ground state energies of nuclei around $^{56}$Ni show that the isovector pair correlations in this region of the nuclide chart correspond to the case of the anharmonic vibration. To describe this situation we should consider a Hamiltonian with a softer $\theta$-dependence of the potential than above. We are planning to do this in a following paper.

\begin{figure}[th]
\centerline{\includegraphics[width=7.3cm]{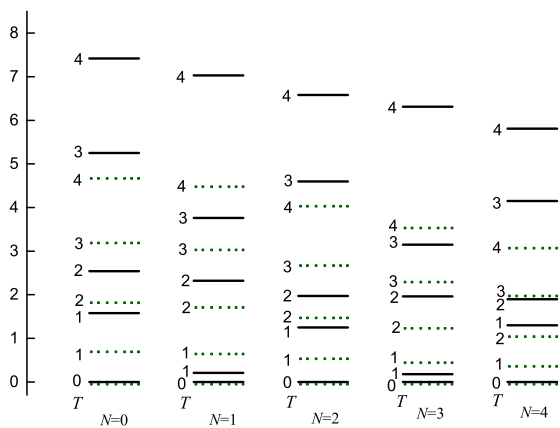}}
\caption{The same as in Fig. 3 but for both even-even and odd-odd nuclei.}
\label{Fig:4}
\end{figure}

 Coming back to description of the energies of the states with $T=0$
we see that calculations based on the model with isovector pair correlations
underestimate binding energies of nuclei with even and equal numbers of protons
and neutrons, i.e. of nuclei which can be presented  as systems of some numbers of
$\alpha$-particles. In our approach properties of the isospin and gauge modes are
determined by the same interaction, namely, by the isovector pairing.
Looks like that these modes are more decoupled and their characteristics are
determined by different components of the nuclear interaction.

We do not analyze in details the results obtained with the $\theta$-independent potentials
since in this case some states with different isospin become degenerate in energy
in contradiction with experiment.

\section{Conclusion}

In the present paper we have performed calculations of the relative energies of the
ground states of nuclei around $^{56}$Ni. The collective Hamiltonian suggested previously
for description of the isovector pair correlations has been used. The calculations have been performed for different variants of the collective potential enabling analytical solutions.

The results of calculations have shown that the isovector pair correlations in nuclei
around $^{56}$Ni are far from being considered as corresponding to the limit of the
static pair correlations. Rather, they can be considered as anharmonic pairing vibrations.

The results of calculations have shown that especially large deviations from the
experimental data are obtained for the ground states of nuclei with even numbers of
$Z$ and $A$, i.e. for nuclei which are systems of some numbers of $\alpha$-particles.

The results of calculations demonstrate a weaker dependence of the relative energies on
isospin compared to experimental data. This indicates that the moment of inertia for isospin rotations in the
model Hamiltonian is significantly larger than the experimental one. This means that
there is some interaction different from isovector pairing which influences the
isospin dependence of the energies.

\section*{Acknowledgements}

The authors express their gratitude to the Russian Foundation for Basic Research (RFBR, grant 20-02-00176) 
and to the Heisenberg-Landau Program for support.

\end{document}